\documentclass[twocolumn,secnumarabic,amssymb, nobibnotes, aps, prd]{revtex4-1}
\usepackage{graphicx,subfigure}
\usepackage{times}
\usepackage{verbatim}
\usepackage{amssymb,amsfonts,amsmath}
\usepackage{algorithm}
\usepackage{algorithmic}
\usepackage{url}
\usepackage{subfigure}
\usepackage{color}

\newcommand{\env}[1]{\texttt{#1}}
\begin{document}

\title{Incremental Versus Optimized Network Design}%

\author{Saeideh Bakhshi}
\email{Saeideh@gatech.edu}
\author{Constantine Dovrolis}
\email{Constantine@gatech.edu}
\affiliation{College of Computing, Georgia Tech}
\begin{abstract}
Even though the problem of network topology design is often studied 
as a ``clean-slate'' optimization, in practice most service-provider 
and enterprise networks are designed incrementally over time.
This evolutionary process is driven by changes in the underlying
parameters and constraints (the ``environment") and it aims to
minimize the modification cost after each change in the environment.
In this paper, we first formulate the incremental design approach
(in three variations),
and compare that with the more traditional optimized design approach 
in which the objective is to minimize the total network cost.
We evaluate the cost overhead and evolvability of incremental design
under two network expansion models (random and gradual),
comparing incremental and optimized networks in terms of cost, topological
similarity, delay and robustness.
We find that even though incremental design has some cost overhead,
that overhead does not increase as the network grows. 
Also, it is less costly to evolve an existing network than to design
it ``from scratch'' as long as the network expansion factor 
is less than a critical value.
\end{abstract}

\maketitle

\section{Introduction}
	\label{sec:intro}
Complex technological systems, such as transportation and
communication networks, manufacturing processes, 
microprocessors and computer operating systems, 
are rarely designed ``from scratch.''  
Instead, they are often subject to an evolutionary process in which
existing designs are incrementally modified every time a new
phase or version of the system is needed.
There are numerous examples and they span every engineering
discipline. To mention one of them, consider a wide-area
communication network that expands over time 
to reach new locations, increasing its capacity depending on 
the offered load, and occasionally providing new services. 

It is remarkable that even though the optimized design of complex 
systems and networks has been studied in depth for several decades, 
the literature rarely considers that the design process is often
incremental (or evolutionary). Instead, it is typically
assumed that the system is designed {\em tabula rasa}, 
in a ``clean-slate'' manner.
The corresponding design problems are typically formulated as 
optimization problems with multiple constraints, none of which
is imposed by an earlier design however. 
In the few previous studies that considered evolving networks
(see Section~\ref{sec:relatedwork}),
the focus was on the design algorithms and the corresponding optimization
problems, rather than to examine the pros and cons of an incremental design 
relative the corresponding optimized design. 

In this paper, we attempt a first investigation of the following
fundamental question: {\em how does an incremental design compare
to an optimized design, when both designs provide the same function?}
Even though we believe that the previous question is relevant to the
study of complex systems in general, we choose to focus on a rather 
narrow design problem and the engineering domain of our expertise,
namely {\em the topological design of communication networks}.
In that context, the objective of the design process is to
create a network that interconnects a given set of locations
under certain reliability and performance constraints, 
aiming to minimize a cost-related objective function.
We further limit (admittedly in a very simplified manner) 
how the environment changes with time:
the set of interconnected network locations expands by one 
or more nodes at each time step.

The previous network topology design problem allows us to examine 
several interesting questions about incremental versus optimized
designs in a precise and quantitative framework.
How can we mathematically formulate the incremental design problem,
and how is that formulation different from the more traditional
optimized design problem?
How does an incremental design compare, in terms of cost, 
topological similarity, performance (delay),
or robustness with the corresponding clean-slate design that provides
the same function?
How costly is it to modify an existing design, relative to 
the cost of re-designing the system from scratch?
When is it better to abandon incremental changes on an existing 
system and start from scratch?
What is the {\em ``price of evolution''}, i.e., the cost overhead of an 
incremental design relative to the corresponding optimized design? 
How different, topology-wise, are two incremental and optimized network designs
that provide the same function?
What is the role of the pace at which the environment changes with time? 
Does the incremental design process perform better when the
environment varies in certain non-random ways?
And finally, how important is it that the incremental design 
process maintains an {\em inventory} of components from 
earlier designs that are not needed in the current design?

In Section~\ref{sec:framework}, we present a 
formulation for the incremental design problem in which the objective
is to {\em minimize the modification cost relative to the existing network.}
We compare that formulation with an optimized design problem in which
the objective is to minimize the total network cost.
In Section~\ref{sec:ring}, we start with a {\em ring design problem.}
In this simpler network structure, we can derive expressions
for the evolvability and cost overhead of the incremental design process. 
We also compare {\em random expansion} with {\em gradual expansion.}
We then switch to the more general {\em mesh network design problem.}
In Section~\ref{sec:mesh}, we describe the optimized
and evolved mesh network algorithms we use in the rest of the paper.
In Section~\ref{sec:basicexpansion}, we compare the incremental 
and optimized design processes under a single-node expansion model
in which one node is added to the network at each new environment. 
In Section~\ref{sec:rho-effect}, we examine a faster expansion
model in which multiple nodes are added simultaneously.
In Section~\ref{sec:robustness}, we compare the robustness
of the optimized and evolved networks in terms of a node centrality metric. 
In Section~\ref{sec:inventory}, we compare three variations
of the incremental design process that differ in how we use 
existing components from previous designs that are not needed
in the current design. 
We review the related work in Section~\ref{sec:relatedwork}
and conclude in Section~\ref{sec:conclusions}.

\section{Framework and Metrics}
	\label{sec:framework}
In this section, we present mathematical formulations for
the clean-slate and incremental design problems. Even though these
formulations are quite general, in the rest of the paper we apply them in 
the context of topology design for communication networks. 
As in any design problem, there is a desired function 
(e.g., construct a communication network to connect a given set of locations),
some design elements (e.g., routers, wide-area links),
certain constraints (e.g., related to reliability or maximum propagation
delay) 
as well as an objective (e.g., minimize the total cost of the required
design elements). 
The design process aims to use the appropriate elements
so that we achieve the desired function, while satisfying the constraints
and meeting a given objective. It is often
assumed that this process is conducted only once in an otherwise static
environment. Here, we consider the case that the design takes place
in a {\em dynamic environment}. In the context of communication networks, 
the network may gradually expand to new locations, 
the cost of design elements may fluctuate,
or the constraints may become more stringent from time to time. 
We consider a discrete-time model and we refer to the $k$'th time epoch as the
{\em k'th environment}. At a given environment $k$, 
all inputs of the design problem are known and constant.

How can we design a communication network in a dynamic environment?
We identify two fundamentally different approaches. 
In the {\em clean-slate approach} we aim {\em to minimize in every environment
the total cost of the network} subject to the given constraints.
We refer to the resulting network in each environment as {\em optimized}. 
In the {\em incremental approach} we aim instead to 
{\em minimize the modification
cost relative to the network of the previous environment}, again subject
to the given constraints; we refer to the resulting network as {\em evolved}.

More rigorously, let $\mathcal{N}(k)$ be the set of 
{\em acceptable networks} at environment $k$, 
i.e., networks that provide the desired function and meet the given 
constraints at environment $k$. 
The cost of a particular network $N\in\mathcal{N}(k)$ is $C(N)$. 
$C(N)$ is the sum of the costs of all design elements in $N$. 
We assume that there are no other costs associated with $N$; for instance, 
there is no monetary cost to compute the design or to interconnect its elements.

In clean-slate design, the objective is to identify an acceptable 
network $N_{opt}(k)$ from the set $\mathcal{N}(k)$ that has the 
minimum cost $C_{opt}(k)$ at environment $k$, 
\begin{equation} 	\label{eq:opt-design}
C_{opt}(k) \equiv C(N_{opt}(k)),\quad 
N_{opt}(k) \equiv \arg\min_{N\in\mathcal{N}(k)}C(N) .
\end{equation}
We refer to $N_{opt}(k)$ as the {\em optimized} network at environment
$k$. If the optimized network is not unique, we break ties with secondary
objectives (for instance, minimize the total number of links). Most
network design problems are computationally intractable (NP-hard),
and so they are often solved heuristically, approximating the previous
optimization objective. This is what we also do in the 
algorithms of Sections~\ref{sec:ring} and \ref{sec:mesh}. For this
reason, we do not refer to $N_{opt}(k)$ as {\em optimal} but as
{\em optimized}. The former would be the actual solution to the
previous problem if we could compute it; the latter is the best solution
we can compute given a certain design heuristic.

In the incremental design approach, on the other hand,
we design the new network $N_{evo}(k)$ based on 
the network $N_{evo}(k-1)$ from the previous environment $k-1$.
We refer to the former as the {\em evolved} network at environment $k$. 
The objective of the incremental design process is to identify an 
acceptable network $N(k)\in\mathcal{N}(k)$ that minimizes the 
{\em modification cost} $C_{mod}(N_{evo}(k-1);N(k))$ between networks 
$N_{evo}(k-1)$ and $N(k)$. 
For simplicity, we denote the previous modification cost as $C_{mod}(k)$. 

To define the modification cost precisely we first have to answer 
the question: what should we do with design elements that
are present in $N_{evo}(k-1)$ but not in $N(k)$? We identify three
options. First, we keep them active in $N(k)$ 
(even though they are not necessary) - 
this is the {\em Ownership} option.
Second, they are removed from $N(k)$ (even though they could be reused
in a future environment) - this is the {\em Leasing} option. 
The third option, that we adopt in most of this paper, is that there 
is an ``inventory'' $I(k)$ of design elements that have been purchased 
prior to environment $k$ but are not used in $N(k)$ - 
this is the {\em Inventory} option.\footnote{In a communication
network, to place a physical link in the inventory means that the
corresponding fiber trunk would not be connected to any switching
equipment, even though it it is already laid of course.}  
We assume that the cost of storing elements in the inventory, 
as well as the cost of moving them in/out of the inventory is zero.
We compare the Ownership and Leasing options with the Inventory
option in Section~\ref{sec:inventory}.

In the presence of an inventory, the modification cost $C_{mod}(k)$
is defined as the cost of new design elements that are needed in $N(k)$
but are not present in $N_{evo}(k-1)$ and they cannot be found in
the inventory $I(k-1)$. 
Formally, $C_{mod}(k)$ is the cost of the design elements in the set 
$S_{mod}(k)$, where 
\begin{equation} 		\label{eq:mod-cost-def}
S_{mod}(k)=N(k)\setminus[N_{evo}(k-1)\cup I(k-1)]
\end{equation}
abusing the notation $N(k)$ to also refer to the set
of design elements in the network $N(k)$. Similarly, the inventory
at environment $k$ includes the design elements that are present
in $N_{evo}(k-1)$ but are not present in $N(k)$, 
\begin{equation} 		\label{eq:inv-def}
I(k)=[N_{evo}(k-1)\cup I(k-1)]\setminus N(k) .
\end{equation}

With the previous definitions, we can now formulate the incremental
design process as: 
\begin{equation} 		\label{eq:evo-design}
C_{evo}(k) \equiv C(N_{evo}(k)),\quad 
N_{evo}(k) \equiv \arg\min_{N\in\mathcal{N}(k)}C_{mod}(k)
\end{equation}
The evolved network $N_{evo}(k)$ may not be unique in general.
Ties are broken by considering a secondary objective: 
if two networks minimize the modification cost, select the network with
the minimum total cost.\footnote{In our computational experiments, 
two modification costs are rarely equal because link costs are based
on distance and they are real numbers.}
As in the case of optimized design, 
we compute the solution of the incremental design problem
with a heuristic described in Sections~\ref{sec:ring} and \ref{sec:mesh}.

The cost of the evolved network can be expressed recursively as ($k\geq1$):
\begin{equation} 		\label{eq:evo-recursive}
C_{evo}(k)=C_{evo}(k-1)+C_{mod}(k) - [C_{inv}(k)-C_{inv}(k-1)]
\end{equation}
where $C_{inv}(k)$ is the cost of all elements in the inventory
at environment $k$. We assume that the initial evolved network 
(and its cost $C_{evo}(0)$) is known,
and that the initial inventory is empty ($C_{inv}(0)=0$).
So, the cost of the evolved network at environment $k$
has increased by the modification cost (new elements that are 
purchased at $k$) minus any potential increase in the inventory's value
at $k$; the last term may be negative.
Note that the cost of the evolved network may decrease at environment $k$
if the modification cost (new design elements) is less than the 
total value of the elements that are moved from the inventory to
the network at time $k$.

Expanding (\ref{eq:evo-recursive}), we can write the cost of the
evolved network as: 
\begin{equation} 		\label{eq:evo-cost}
C_{evo}(k)=C_{evo}(0)+\sum_{i=1}^{k}C_{mod}(i)-C_{inv}(k).
\end{equation}
Thus, the cost of the evolved network at environment $k$ is the
cost of the initial network plus the cost of all design elements that
were purchased in the last $k$ environments, minus anything that
remains in the inventory at time $k$.

\subsection{Metrics}
We now introduce four metrics to compare a sequence of
optimized and evolved networks. We also introduce two specific 
models of dynamic environment we consider in this paper, and quantify
the rate at which the environment changes with time.

First, the {\em cost overhead} $v(k)$ of the evolved design $N_{evo}(k)$
relative to the corresponding optimized design $N_{opt}(k)$ at environment
$k$ is: 
\begin{equation} 		\label{eq:costoverhread}
v(k)=\frac{C_{evo}(k)}{C_{opt}(k)}-1\geq0
\end{equation}
where the inequality is expected from the definition of $C_{opt}(k)$.
\footnote{The reader should note that the cost overhead metric is different 
than the well-known {\em approximation ratio}. The latter examines
the worse-case solution produced by an algorithm relative to the optimal 
solution of a given problem. The cost overhead compares the solutions 
(costs)
of two algorithms that aim to minimize two {\em different} optimization 
objectives.}
What is more important however is whether the cost overhead of the
incremental design process gradually increases, i.e., whether the
evolved networks become increasingly more expensive compared to the
corresponding optimized networks. If that is the case, the incremental
design process would diverge over the long-term towards extremely
inefficient designs.

Second, the {\em evolvability} $e(k)$ is defined as: 
\begin{equation} 		\label{eq:evolvability}
e(k)=1-\frac{C_{mod}(k)}{C_{opt}(k)}\leq1 .
\end{equation}
The evolvability represents the cost of modifying the evolved network
from environment $k-1$ to $k$, relative to the cost of redesigning
the network ``from scratch'' at time $k$. High evolvability, close
to 1, means that it is much less expensive to modify the existing
network than to re-design a new network. On the other hand, when the
evolvability becomes zero or negative,
it is beneficial to stop the incremental design process and design
a new optimized network, i.e., clean-slate design {}``beats'' evolution
in that case.

Third, the {\em inventory overhead} $r(k)$ is defined as: 
\begin{equation} 		\label{eq:inventory-overhead}
r(k)=\frac{C_{inv}(k)}{C_{evo}(k)}\geq0 .
\end{equation}
The inventory overhead quantifies the cost of design elements that
have been previously purchased but are now left unused, relative to
the current cost of the evolved network. An incremental design process
that leads to a gradually increasing inventory overhead would be 
inefficient in terms of its cumulative cost over time.

Fourth, the {\em topological similarity} $t(k)$ between the optimized
$N_{opt}(k)$ and evolved $N_{evo}(k)$ networks is defined as the
{\em Jaccard similarity coefficient} of the two corresponding adjacency
matrices. In other words, $t(k)$ is the fraction of distinct links
in either network that are present in both networks. Even though the
two network design approaches we consider are different in terms of objective
and design method, we are interested to know how different the resulting
networks are, structure-wise, over time. 

\subsection{Expansion models}
We consider a specific way in which the environment changes with time: 
{\em expansion.} Specifically,
the set of locations that the network has to interconnect at any
time $k$ is increasing with $k$. This is probably the most natural way the 
environment can change with time in the context of communication networks. 

In the simplest form of expansion, the network size increases by 
only one node at each environment; we refer to this as 
{\em single-node expansion}.

We also consider a {\em multi-node expansion} scenario in which the network size
increases once by a multiplicative factor $\rho$,  which we refer to as
{\em expansion factor}. Specifically, if the network size increases
from $n$ nodes to $n+m$ nodes, the expansion factor $\rho$ is 
\begin{equation} 			\label{eq:rho-def}
\rho=\frac{n+m}{n}\geq1 .
\end{equation}

We also compare two expansion models: {\em random} and {\em gradual}.
Suppose that the set of all possible locations is $\cal{L}$
and that the network expands at time $k$ to $X$ new locations. 
In random expansion, the $X$ new locations are selected randomly 
from $\cal{L}$.
In gradual expansion, we select iteratively each of the $X$ new locations
from $\cal{L}$ so that it is the closest location to either any of the existing
nodes in network $N(k-1)$ or to any of the new locations we have just added 
in the network. 

The random and gradual expansion models represent two significantly
different models in which the environment changes with time.
In random expansion, the new locations can be anywhere and so it may be 
costly for the incremental design to adjust the previous network
with only minor modifications.
In gradual expansion, the environment changes in an ``evolution-friendly''
manner because the new locations are as close as possible to the existing 
network.

\section{Ring Networks}
	\label{sec:ring}
In this section, we compare the optimized and incremental
design approaches in the context of a ring topology.
Ring networks are widely used mostly in metropolitan-area networks,
as they are robust to single-node failures (two node-disjoint paths
exist between any pair of nodes) and they are typically less costly
than mesh networks \cite{man-ring}.
For our purposes, the ring topology has two additional features.
First, we can use an existing software package (Concorde) \cite{concorde}
that computes excellent approximations to the optimal ring design problem.
Second, we can leverage existing analytical results to derive
asymptotic expressions for the evolvability and cost overhead of incremental
ring design under random and gradual expansion.

\subsection{Optimized versus incremental ring design}
We assume that all potential nodes of the expanding 
ring network are located in a bounded plane region.
Further, we assume that the cost of a network 
is equal to the sum of its link costs, and the cost of each link
is proportional to its length. 
So, the minimum-cost ring design problem is equivalent to the
NP-Hard Traveling Salesman Problem (TSP) \cite{TSP-hard}.
We rely on a TSP-solving software package called 
Concorde \cite{concorde}, which is based on a branch-and-cut algorithm.
Concorde's TSP solver has computed the {\em optimal} solutions 
to 106 of the 110 {\em TSPLIB} problem instances; the largest of them
has 15,112 nodes. 

Further, in the case of ring design we can use a simple asymptotic
expression for the length of the optimal TSP tour. 
Specifically, Beardwood et al. proved that 
{\em ``the length of the shortest closed path through $n$ points in a bounded 
plane region of area $A$ is almost always asymptotically proportional to 
$\sqrt{A \times n }$ for large $n$''} \cite{bearwood}.
In our context, the length of the TSP tour is equal to the
cost of the optimized ring.

In the case of random expansion, the $n$ nodes of the ring can 
be anywhere in the given region and so the area $A$ does not depend on $n$.
Thus, the cost of the optimized ring increases as 
\begin{equation} 		\label{eq:Ring-C_opt}
C_{opt}^{rnd}(n) \sim \sqrt{A \times n} \sim \sqrt{n}
\end{equation}
where the notation $x \sim f(n)$ means that $x$ tends to become 
proportional to $f(n)$ as $n$ increases. 

In the case of gradual expansion, the area in which the $n$
ring nodes are located increases with $n$. If we assume that all
possible nodes are uniformly distributed with point density 
$\sigma_{grd}$  in a bounded
plane region, then the area $A$ in which $n$ ring nodes are
located increases proportionally with $n$ ($n = A \, \sigma_{grd}$).
Thus, 
\begin{equation} 		\label{eq:Ring-C_opt-grd}
C_{opt}^{grd}(n) \sim \sqrt{\sigma_{grd} \times n^2} \sim  n .
\end{equation}
\begin{figure}[ht]
\begin{center}
\subfigure[ ]{
\includegraphics[width=0.18\textwidth]{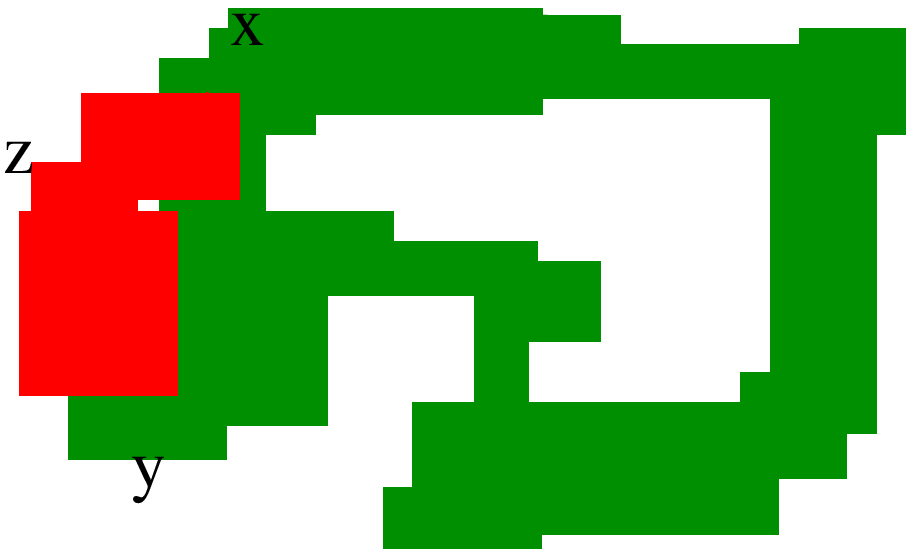}
\label{fig:one-node}
}
\subfigure[ ]{
\includegraphics[width=0.22\textwidth]{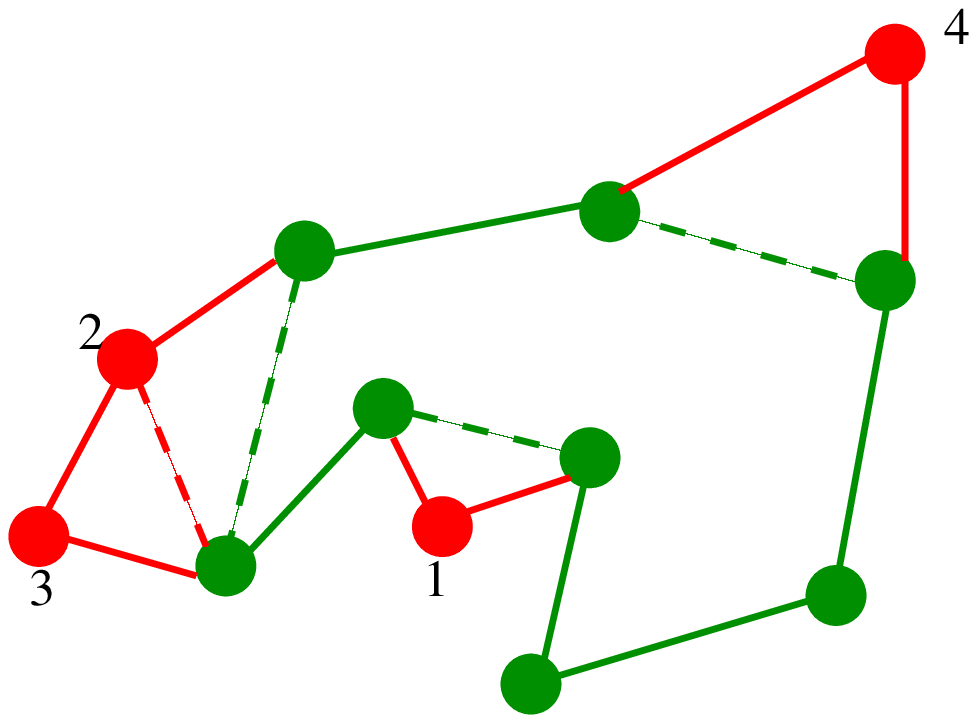}
\label{fig:multinodes}
}
\caption{Connecting new nodes to an existing ring: (a) single-node expansion, 
(b) multi-node expansion (we show the order in which nodes are connected).}
\end{center}
\label{fig:Ring connect-new-nodes}
\end{figure}
In the incremental ring design process, we can compute
the minimum modification cost under single-node expansion.
Suppose that the existing ring $N_{evo}(k-1)$ has size $n$ and we add a
single extra node $z$ at time $k$.
The minimum modification cost will result if we connect $z$ to
two adjacent nodes $x$ and $y$ of $N_{evo}(k-1)$, such that
\begin{equation} 		\label{eq:Ring-Cmod}
C_{mod}(k)=min_{(x,y)\in N_{evo}(k-1)} (||z-x||+||z-y||)
\end{equation}
and then removing the edge $(x,y)$.\footnote{We could 
move that edge to the inventory,
but in the case of ring design the inventory is never used. So,
in this section, the Inventory and Leasing options are equivalent.}
This process is illustrated in Figure~\ref{fig:one-node}.
Note that there is no other way to connect $z$ to $N_{evo}(k-1)$
so that the resulting network is still a ring
but with lower modification cost.

In the case of multi-node expansion, we use an 
iterative heuristic that aims to minimize the modification cost.
Suppose that the existing ring $N_{evo}(k-1)$ has size $n$ and we add 
a set $Z$ of more than one new nodes at time $k$.
In each iteration, we select the node $z$ from $Z$ that
minimizes the expression (\ref{eq:Ring-Cmod}), 
connect $z$ to the existing ring as we do in the case of single-node expansion,
and then move $z$ from $Z$ to the set of nodes in $N_{evo}(k-1)$. 
This process is illustrated in Figure~\ref{fig:multinodes}.
Note that this greedy heuristic may be sub-optimal in minimizing 
the modification cost $C_{mod}(k)$ - an exhaustive search for the
optimal solution however can be prohibitively slow. 

\begin{figure*}[hbtp]
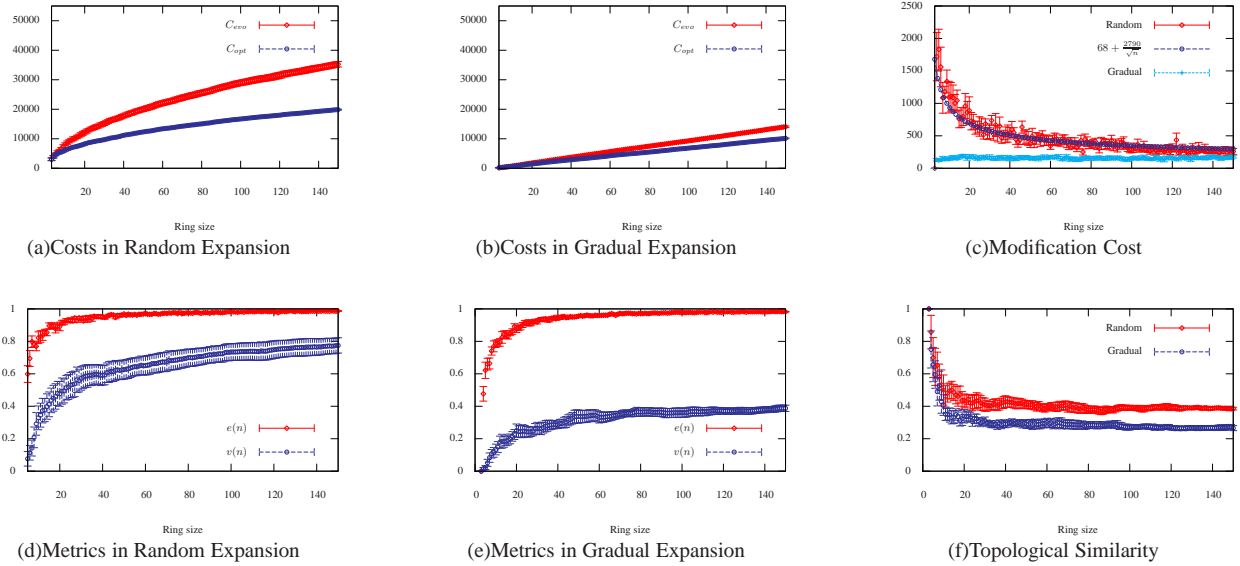

\begin{center}
\begin{tabular}[t]{cccc}
\subfigure[Costs in Random Expansion]{
\scalebox{0.45}{\input{Figures/ring/costsEOR.tex}}
\label{fig:ring random costs}
}
\subfigure[Costs in Gradual Expansion]{
 \scalebox{0.45}{\input{Figures/ring/costsEOG.tex}}
\label{fig:ring gradual costs}
}
\subfigure[Modification Cost]{
\scalebox{0.45}{\input{Figures/ring/Cmod.tex}}
\label{fig:ring Cmod}
}\\
\subfigure[Metrics in Random Expansion]{
\scalebox{0.45}{\input{Figures/ring/metricsR.tex}}
\label{fig:ring random metrics}
}
\subfigure[Metrics in Gradual Expansion]{
\scalebox{0.45}{\input{Figures/ring/metricsG.tex}}
\label{fig:ring gradual metrics}
}
\subfigure[Topological Similarity]{
\scalebox{0.45}{\input{Figures/ring/t.tex}}
\label{fig:ring t}
}
\end{tabular}
\caption{Results for ring network under random and gradual single-node expansion.}
\label{ring}
\end{center}
\end{figure*}

\subsection{Random single-node expansion}
To simplify the notation, instead of referring to an environment 
$k$, we simply refer to the ring size $n$. 
So, instead of writing $C_{mod}(k)$ we write $C_{mod}(n)$,
which refers to the modification cost when the ring expands 
to $n$ nodes. 

To compute the modification cost under random expansion, we rely 
on a result for the nearest neighbor problem:
given a set $S$ of $n$ points and a new point $z$ find the closest neighbor 
of $z$ in $S$.
When $n$ points reside in a two-dimensional region of size $A$ 
with point density $\sigma_{rnd}$, the expected value 
of the nearest neighbor distance is 
$1/\sqrt{\sigma_{rnd}}=1/\sqrt{n/A}$ \cite{nearest-neighbor}.
The modification cost in Equation (\ref{eq:Ring-Cmod})
can be approximated as twice the distance between the new node 
and the nearest node in the ring.
The expected value of the latter is $1/\sqrt{n/A}$ because the $n$
ring nodes are randomly placed in the region $A$. 
So, based on the previous approximation,  the modification cost $C_{mod}(n)$ 
also scales as the mean nearest neighbor distance,
\begin{equation}	\label{eq:rnd-cmod}
C_{mod}^{rnd}(n) \sim \frac{1}{\sqrt{n}}.
\end{equation}
We have confirmed the validity of the previous approximation with 
computational results (see Figure~\ref{fig:ring Cmod}). 

Using Equation (\ref{eq:evo-cost}), and considering that the
incremental ring design process does not use the inventory, 
we see that the cost of the evolved network has the same
scaling behavior as the cost of the optimized network, 
\begin{equation} 	\label{eq:Ring-c_evo}
C_{evo}^{rnd}(n) = \sum_{i=2}^n C_{mod}^{rnd}(i) \sim \sqrt{n}.
\end{equation}
We can now derive asymptotic expressions for the evolvability
and cost overhead under random expansion.
From (\ref{eq:Ring-C_opt}) and (\ref{eq:rnd-cmod}), it follows that
\begin{equation}	\label{eq:rnd-evolvability}
1-e^{rnd}(n) \sim \frac{1}{n}.
\end{equation}
Thus, the evolvability under single-node expansion converges to one, 
i.e., for large rings, the modification cost is practically zero
compared to the cost of designing a new optimized ring.
Also, we expect that the cost overhead 
will be practically constant for large values of $n$ 
because both the evolved and
optimized network costs scale as $\sqrt{n}$,
\begin{equation}
v^{rnd}(n) \sim \mbox{constant}.
\end{equation}
Thus, the evolved ring does {\em not} become increasingly more 
expensive relative to the optimized ring under random single-node expansion.
The exact value of the cost overhead depends on the placement of
the nodes, the order in which they are added to the network,
and the initial ring we start from. 

We have confirmed these asymptotic expressions with computational
experiments in which the optimized ring is designed using Concorde
and the evolved ring is designed based on Equation \ref{eq:rnd-cmod}. 
Figure~\ref{fig:ring random costs} shows the optimized and evolved 
network costs
(with 90\% confidence intervals for the empirical results after 20 runs),
Figure~\ref{fig:ring Cmod} shows the modification cost, while 
Figure~\ref{fig:ring random metrics} shows the evolvability and cost overhead.
In the last graph, note that the cost overhead appears to 
increase with the ring size - we have confirmed that for larger
ring sizes (larger than approximately 200 nodes) 
the cost overhead converges to about 0.8.

\subsection{Gradual single-node expansion}
In this case, a new node is selected
among all potential locations as the closest location to any existing ring node.
We can rely again on the nearest neighbor problem to estimate the modification
cost. Suppose that all potential new locations are uniformly distributed
in the given area with density $\sigma_{grd}$.
The expected value of the distance between a node in the current ring
(of size $n$) and the nearest potential location is $1/\sqrt{\sigma_{grd}}$,
which does not depend on $n$.
The modification cost can be again approximated by twice the previous distance,
and so it should not increase with $n$, at least for large rings,
\begin{equation}	\label{eq:grd-cmod}
C_{mod}^{grd}(n) \sim \mbox{constant}.
\end{equation}
Using Equation (\ref{eq:evo-cost}), and considering that the
incremental ring design process does not use an inventory, 
we see that the cost of the evolved network under gradual
expansion scales linearly with $n$, 
\begin{equation} 	\label{eq:Ring-c_evo-grd}
C_{evo}^{grd}(n) = \sum_{i=2}^n C_{mod}^{grd}(i) \sim n.
\end{equation}
So, the evolvability under gradual expansion scales
as in the case of random expansion
\begin{equation}	\label{eq:grd-evolvability}
1-e^{grd}(n) \sim \frac{1}{n}.
\end{equation}
The cost overhead does not increase with $n$ for large values of $n$,
as in the case of random expansion,
\begin{equation}
v^{grd}(n) \sim \mbox{constant}
\end{equation}
based on Equations (\ref{eq:Ring-c_evo-grd}) and (\ref{eq:Ring-C_opt-grd}).

The difference between random and gradual single-node expansion
becomes evident in the computational results
(see Figure~\ref{ring}).
For the same ring size, the evolved and optimized
costs are lower under gradual expansion,  
the modification cost is practically constant 
(and lower than under random expansion), 
the evolvability is not significantly different between the
two expansion models,
while the cost overhead is significantly lower under gradual
compared to random expansion.
In other words, when the ring expands in a gradual manner,
we expect that the evolved network will be closer, in terms of cost, 
to the optimized network compared to random expansion. 

\subsection{Effect of expansion factor $\rho$}
Suppose that we add $m$ new nodes to a ring of size $n$
so that the resulting network is a ring with $n+m$ nodes.
The expansion factor is $\rho=\frac{n+m}{n} > 1$.
What can we expect about the evolvability and cost overhead
of the incremental design process as functions of $\rho$? 

If $m \ll n$ (i.e., $\rho$ is close to one), 
we can rely on the following simple approximation.
Under random expansion, the modification cost $C_{mod}^{rnd}(n;m)$ 
when we add $m$ new nodes will be approximately $m$ times larger than the
modification cost $C_{mod}^{rnd}(n)$ when we add a single new
node at a ring of size $n$. So,
\begin{equation}
C_{mod}^{rnd}(n;m) \approx m \, C_{mod}^{rnd}(n) \sim m/\sqrt{n}.
\end{equation}
On the other hand, the cost of an optimal ring with $n+m$
nodes will be 
\begin{equation}
C_{opt}^{rnd}(n;m) \sim \sqrt{n+m}.
\end{equation}
Thus, the evolvability under random expansion scales as
\begin{equation}
1-e^{rnd}(n;m)=\frac{C_{mod}^{rnd}(n;m)}{C_{opt}^{rnd}(n;m)}
\sim \frac{m}{\sqrt{n(n+m)}}
\end{equation}
that can be written as 
\begin{equation}	\label{eq:rnd-evo-rho}
1-e^{rnd}(n;m) \sim \frac{\rho-1}{\sqrt{\rho}}.
\end{equation}
As $\rho$ increases the evolvability decreases 
and there is a critical expansion factor value
$\hat{\rho}$ at which the evolvability becomes zero.  
For larger expansions than $\hat{\rho}$, 
it is better to abandon the existing
network and redesign the ring in a clean-slate manner.

\begin{figure}[ht]
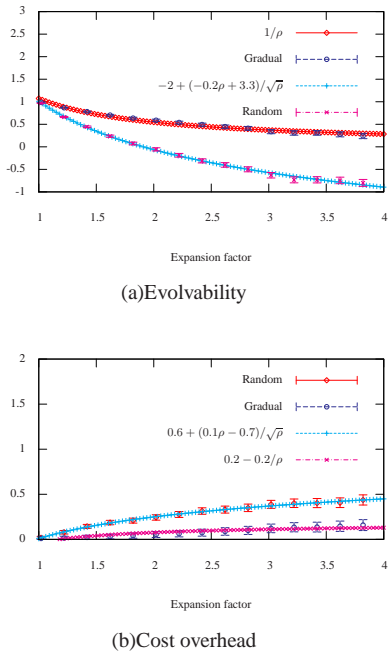

\begin{center}
\subfigure[Evolvability]{
\scalebox{0.5}{\input{Figures/ringPace/Epace.tex}}
\label{fig:epace}
}\\
\subfigure[Cost overhead]{
\scalebox{0.5}{\input{Figures/ringPace/Vpace.tex}}
\label{fig:vpace}
}
\caption{Effect of expansion factor under random and gradual expansion.}
\end{center}
\label{fig:ring pace}
\end{figure}

In the case of gradual expansion, the modification cost $C_{mod}^{grd}(n;m)$ 
will again be roughly $m$ times larger than the
modification cost $C_{mod}^{grd}(n)$ when we add only one node, 
\begin{equation}
C_{mod}^{grd}(n;m) \approx m \, C_{mod}^{grd}(n) \sim m \times \mbox{constant} 
\end{equation}
while the cost of an optimal ring with $n+m$ nodes is 
\begin{equation}
C_{opt}^{grd}(n;m) \sim n+m.
\end{equation}
Thus, the evolvability under gradual expansion scales as
\begin{equation}
1-e^{grd}(n;m)=\frac{C_{mod}^{grd}(n;m)}{C_{opt}^{grd}(n;m)}
\sim \frac{m}{n+m}
\end{equation}
that can be written as 
\begin{equation}	\label{eq:grd-evo-rho}
e^{grd}(n;m) \sim \frac{1}{\rho}.
\end{equation}
Again, as $\rho$ increases the evolvability decreases, but
more slowly than under random expansion. 
It is also important that a critical expansion factor
at which the evolvability becomes zero may {\em not} exist under
gradual expansion.
Equation (\ref{eq:grd-evo-rho}) was derived
assuming that $\rho$ is close to one, and so we cannot  rely
on that expression to prove that the evolvability is always positive. 
Computational results, however, indicate that this may be the case
under gradual expansion (see Figure~\ref{fig:epace}). 

To derive the cost overhead under multi-node expansion, we 
have to make two additional assumptions.
First, the evolved network at size $n$ is the same with 
the optimized network at size $n$ (i.e., $C_{evo}(n)=C_{opt}(n)$). 
Second, because $m \ll n$, the cost of the optimized
network at size $n$ is approximately equal to the cost
of the optimized network at size $n+m$ 
(i.e., $C_{opt}(n) \approx C_{opt}(n+m)$). 
Then, under both random and gradual expansion, we can write that
\begin{equation} 	
C_{evo}(n+m) = C_{evo}(n) + C_{mod}(n;m) \approx
C_{opt}(n+m) + m \, C_{mod}(n) 
\end{equation}
and so the cost overhead is 
\begin{equation} 	
v(\rho) = \frac{C_{evo}(n+m)}{C_{opt}(n+m)}-1 \approx m \, 
\frac{C_{mod}(n)}{C_{opt}(n)}.
\end{equation}
Thus, under random expansion, the cost overhead scales as  
\begin{equation}	\label{eq:rnd-ovh-rho}
v^{rnd}(\rho) \sim \frac{\rho-1}{\sqrt{\rho}}
\end{equation}
while under gradual expansion it scales as
\begin{equation}	\label{eq:grd-ovh-rho}
v^{grd}(\rho) \sim 1- \frac{1}{\rho}.
\end{equation}
The previous scaling expressions are derived
assuming that $\rho$ is close to one, but computational
results confirm that they are quite accurate when $\rho$ is as high as four
(see Figure~\ref{fig:vpace}). 

We conducted computational experiments in which an initial
ring (of size 50) is increased, under random or gradual multi-node expansion,
by different factors $\rho$.
Figure~\ref{fig:ring pace}  shows the evolvability and cost overhead 
as functions of $\rho$.
The previous scaling expressions, which are derived assuming that $n$
is large and $\rho$ is close to one, are actually in close agreement 
with the numerical results when $\rho$ is as high as four. 
Under random expansion, the critical expansion factor 
is $\hat{\rho}\approx 2$, while the corresponding cost overhead 
is approximately 25\%, meaning that even though it is then beneficial 
to re-design the network from scratch, the cost overhead of the evolved 
network is still quite low. 
Under gradual expansion, the evolvability does not become negative,
at least in the given range of $\rho$, and the cost overhead is significantly
lower than under random expansion.

\section{Mesh Network Design}
	\label{sec:mesh}
We now describe the algorithms that we use to design optimized
and evolved mesh networks. The main difference between mesh and ring
topologies, at least in the context of this work, 
is that the former has an additional design constraint that is 
related to the delay of each path. 
Even though
ring topologies are common in metropolitan networks, mesh networks are
the norm in wide-area service provider backbones and enterprise networks.
A ring topology is not appropriate in that context because it can
lead to unacceptably large delays.

Specifically, 
the reliability constraint
is that every pair of networks should be connected through
at least {\em two node-disjoint paths}, the primary and the secondary. 
The primary path is the path with the minimum propagation delay, 
i.e., the shortest path when each link cost is equal to the link's 
propagation delay.\footnote{We assume that the propagation delay of a 
link is proportional to its straight-line length on the Euclidean plane.} 
The secondary path is also the shortest path, but after we have removed
the nodes that participate in the primary path (except the source
and destination).
The propagation delay in both paths should be less than $D$. 
We say that a network is {\em acceptable} if it meets the previous
reliability and delay constraints for every pair of nodes. 
Note that depending on the distances between nodes and the bound $D$
an acceptable network may not exist. 

Given a set of locations, 
we only consider the cost of links (edges). The routers (nodes)
would introduce the same cost in all networks as long as we have one
router at each location. 
As in the case of ring networks, we assume that the cost of a link is 
proportional to its length. In other words, we focus on the cost of installing
say a fiber optic trunk between two remote locations. That cost is
typically much larger than the cost of the router interfaces for a 
link, it does not depend on the capacity of the link, 
and it increases roughly linearly with distance 
(at least for transcontinental links).

Because the problem of minimum-cost topological design for mesh
networks with reliability and delay constraints is NP-Hard
\cite{network-design}, we rely on heuristics. 
Even though
there are some approximation bounds for special networks, mostly trees,
we are not aware of such bounds and approximation algorithms for general 
mesh networks under the previous design constraints.
Our objective is not to study network design algorithms but to compare
optimized with evolved designs, and so we use two rather simple algorithms
referred to as {\em OPT} and {\em EVO}. Both algorithms are
probabilistic and iterative. The two algorithms are quite similar
in terms of how they add and remove links, but they differ 
in their objective functions. Additionally, the {\em EVO} algorithm
makes use of links stored in an inventory.

In the {\em OPT} algorithm the objective function
is to minimize the total network cost. That is computed as the sum
of all link costs. In the {\em EVO} algorithm, the objective
is to minimize the modification cost relative to the previous
network, reusing any links that may exist in the inventory.
In {\em EVO}, the modification cost is the sum of the costs for
all {\em new links} that are needed in the evolved network.
That cost does not include the cost of any existing
links in the previous network, or any links that are moved from the
inventory to the new network.

In both algorithms, we use the same stopping criterion. Because the
algorithms are probabilistic, each iteration may result in a different 
acceptable network (if such a network exists). If the new network
is {\em not} better, in terms of the optimization objective of each algorithm,
than the best network that has been computed up to that point,
we move to the next iteration. The algorithms terminate if we cannot
improve the optimization objective of each algorithm for a number
(10) of successive iterations.
It should be noted that we have also experimented with several other 
algorithms. The aforementioned performed consistently better
in terms of minimizing the two corresponding objective functions
than any other heuristic we experimented with. 
So, even though we cannot claim that the following algorithms
are optimal or that they have a certain approximation ratio, 
we are certain that they do not produce	``bad'' solutions either.

\subsection{The OPT algorithm}
Algorithm ~\ref{algo:optimized} describes a single iteration of the
{\em OPT} design process. Each iteration aims to find an acceptable network.
At the end we choose 
the network with minimum total cost among all the designed acceptable networks.
Each iteration has two phases. In the first phase, the
algorithm adds links probabilistically, in order of increasing cost,
until an acceptable network is computed. 
In the second phase, the
algorithm attempts to remove as many existing links as possible, in
order of decreasing cost, as long as the network remains acceptable.
That stage is also probabilistic.
\algsetup{indent=1em}
\begin{figure}[htp]
\caption{OPT-design-iteration}
\label{algo:optimized}
\begin{algorithmic}[1]
\REQUIRE Initial network $N_{init}$: Concorde ring connecting all nodes 
\STATE FoundAcceptable = false
\STATE $N_{OPT}'$ =  graph complement of $N_{OPT}$
\STATE LinkstoAdd = all the links in $N_{OPT}'$ \\
\COMMENT {Link addition phase}
\WHILE {LinkstoAdd is {\em not} empty and {\em not} FoundAcceptable}
  \STATE  $s$ = shortest link in LinkstoAdd
  \STATE Add $s$ to $N_{OPT}$ with probability $p_{add}$
  \STATE remove $s$ from LinkstoAdd
  \IF{$N_{OPT}$ is acceptable}
    \STATE FoundAcceptable = true
  \ENDIF
\ENDWHILE \\
\COMMENT {Link deletion phase}
\STATE LinkstoRemove = all the links in $N_{OPT}$
\WHILE {LinkstoRemove is {\em not} empty and foundAcceptable}
    \STATE  $s$ = longest link in $N_{OPT}$
    \STATE Examine if $N_{OPT}$ would remain acceptable if $s$ is removed
    \STATE If so, remove $s$ with probability $p_{del}$
    \STATE remove $s$ from LinkstoRemove
\ENDWHILE
\RETURN Optimized network $N_{OPT}$ and FoundAcceptable
\end{algorithmic}
\end{figure}
The previous process starts from 
an optimized ring that interconnects all given locations, computed
using the Concorde TSP solver (see Section~\ref{sec:ring}).

We found empirically that the two probabilities $p_{add}$ and
$p_{del}$ do not have a strong impact on the resulting minimum network
cost, as long as they are between 0.8 and 1; we use $p_{add}$=$p_{del}$=0.9.
 
\subsection{The EVO algorithm}
Algorithm~\ref{algo:evolved} describes a single iteration of the
{\em EVO} design process. The algorithm is similar to {\em OPT}
in the way it adds and deletes links, but with three important
differences. 
\algsetup{indent=1em}
\begin{figure}[htp]
\caption{EVO-design-iteration}
\label{algo:evolved}
\begin{algorithmic}[1]
\REQUIRE Previous evolved network $N_{prev}$, previous inventory $I_{prev}$, and  set of new nodes. 
\STATE FoundAcceptable = false
\STATE $N_{EVO}=N_{prev}$
\STATE add all the links from $I_{prev}$ to $N_{EVO}$
\STATE connect new nodes to $N_{EVO}$ (see Figure~1)
\STATE $N_{EVO}'$ =  graph complement of $N_{EVO}$
\STATE LinkstoAdd = all the links in $N_{EVO}'$ \\
\COMMENT {Link addition phase}
\WHILE {LinkstoAdd is {\em not} empty and {\em not} FoundAcceptable}
  \STATE  $s$ = shortest link in LinkstoAdd
  \STATE Add $s$ to $N_{EVO}$ with probability $p_{add}$
  \STATE remove $s$ from LinkstoAdd
  \IF{$N_{EVO}$ is acceptable}
    \STATE FoundAcceptable = true
  \ENDIF
\ENDWHILE \\
\COMMENT {Link deletion phase, first removing links from newly added links, then the existing links}
\STATE LinkstoRemove = all the links in $N_{EVO}-(N_{prev} \bigcup I_{prev})$
\WHILE {LinkstoRemove is {\em not} empty and FoundAcceptable}
    \STATE  $s$ = longest link in linkstoRemove
    \STATE Examine if $N_{EVO}$ would remain acceptable if $s$ is removed
    \STATE If so, remove $s$ with probability $p_{del}$
    \STATE remove $s$ from LinkstoRemove
\ENDWHILE
\STATE LinkstoRemove = all the links in $N_{prev} \bigcup I_{prev}$
\WHILE {LinkstoRemove is {\em not} empty and FoundAcceptable}
    \STATE  $s$ = longest link in LinkstoRemove
    \STATE Examine if $N_{EVO}$ would remain acceptable if $s$ is removed
    \STATE If so, remove $s$ with probability $p_{del}$
    \STATE remove $s$ from LinkstoRemove and add it to $I_{curr}$
\ENDWHILE
\RETURN {Evolved network $N_{EVO}$, current inventory $I_{curr}$ and FoundAcceptable}
\end{algorithmic}
\end{figure}
\begin{figure*}[ht]
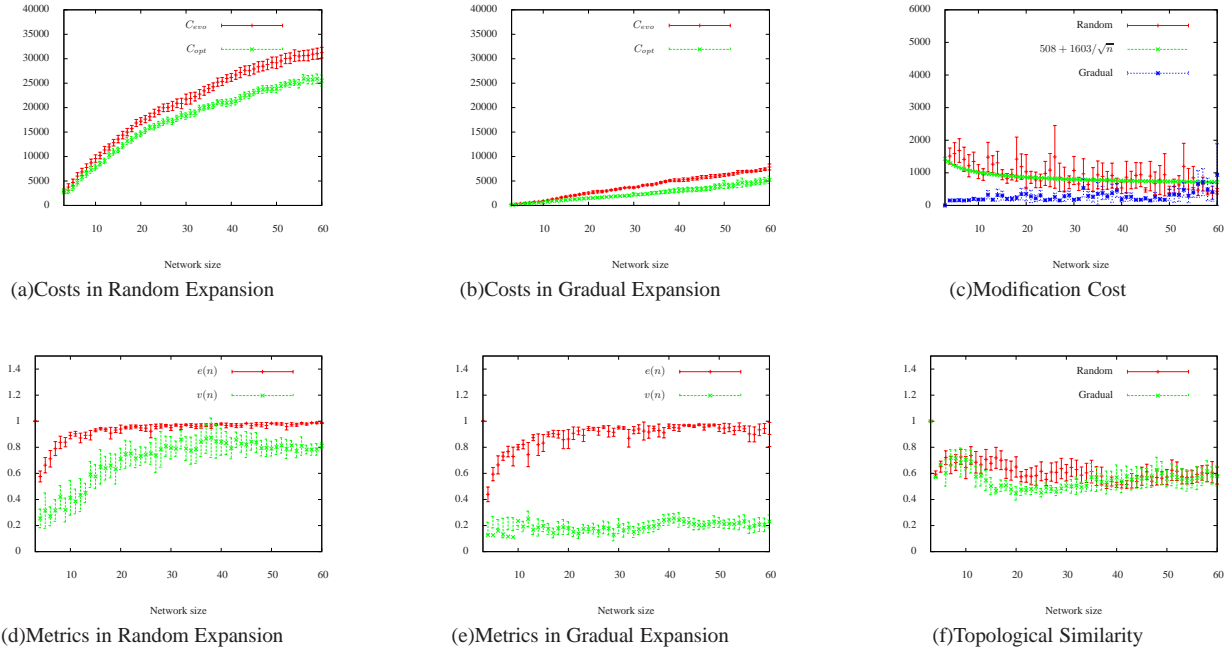

\begin{center}
\begin{tabular}[t]{ccc}
\subfigure[Costs in Random Expansion]{
\scalebox{0.45}{\input{Figures/basicExpansion/RCOSTS.tex}}
\label{fig:basic r Costs}
}
\subfigure[Costs in Gradual Expansion]{
\scalebox{0.45}{\input{Figures/basicExpansion/GCOSTS.tex}}
\label{fig:basic g Costs}
}
\subfigure[Modification Cost]{
\scalebox{0.45}{\input{Figures/basicExpansion/CMODRG.tex}}
\label{fig:basic rg Cmod}
}\\
\subfigure[Metrics in Random Expansion]{
\scalebox{0.45}{\input{Figures/basicExpansion/RMETRICS.tex}}
\label{fig:basic r metrics}
}
\subfigure[Metrics in Gradual Expansion]{
\scalebox{0.45}{\input{Figures/basicExpansion/GMETRICS.tex}}
\label{fig:basic g metrics}
}
\subfigure[Topological Similarity]{
\scalebox{0.45}{\input{Figures/basicExpansion/TRG.tex}}
\label{fig:basic rg T}
}
\end{tabular}
\caption{Results of mesh networks under random and gradual single-node expansion.}
\label{gradual expansion}
\end{center}
\end{figure*}
First, {\em EVO} connects the given set of new nodes to the
previous network using the iterative process that was also used
in Section~\ref{sec:ring}; recall that that algorithm aims to connect each 
new node to the existing network introducing the lowest modification cost.

Second, {\em EVO} attempts to reuse links from the inventory as much
as possible so that it minimizes the new links
that we need to acquire at this environment. Note that the latter
are the only links that contribute to the modification cost.

Third, {\em EVO} has two link deletion phases. It first removes
(probabilistically) new links that are not necessary in order
of decreasing cost. Any link deletions in this phase reduce the
modification cost. Then, it moves 
(again, probabilistically and in order of decreasing cost)
existing links that are not necessary into the inventory.
Any link deletions in this phase do not reduce the modification
cost, but they reduce the cost of the evolved network.
Note that the latter is a secondary objective, and it is 
pursued only after we have reduced the modification cost as much as possible.

We use the same values for $p_{add}$ and $p_{del}$ as in {\em OPT}.

\section{Single-Node Mesh Expansion}
	\label{sec:basicexpansion}
\begin{figure*}[ht]
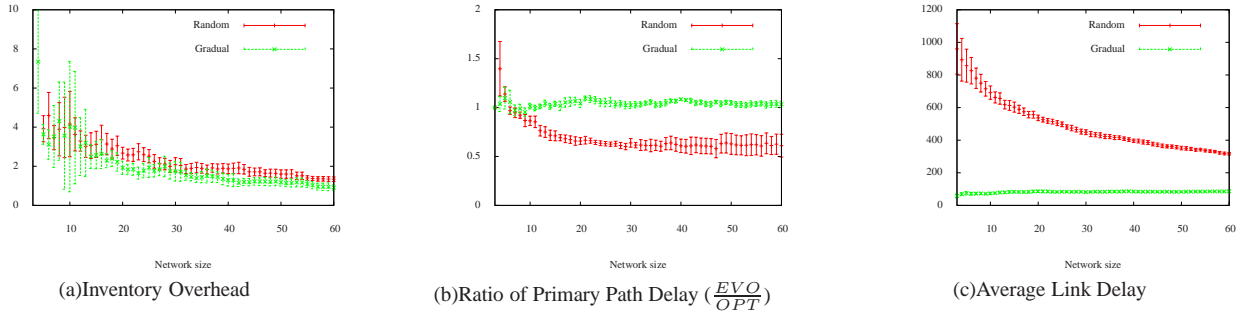

\begin{center}
\begin{tabular}[t]{cccc}
\subfigure[Inventory Overhead]{
\scalebox{0.45}{\input{Figures/basicExpansion/RRG.tex}}
\label{fig:basic rg R}
}
\subfigure[Ratio of Primary Path Delay ($\frac{EVO}{OPT}$)]{
 \scalebox{0.45}{\input{Figures/basicExpansion/R2RG.tex}}
\label{fig:basic rg pathlength}
}
\subfigure[Average Link Delay]{
\scalebox{0.45}{\input{Figures/basicExpansion/R4RG.tex}}
\label{fig:basic rg average linklength}
}
\end{tabular}
\caption{Three properties under single-node expansion.}
\label{random expansion topology}
\end{center}
\end{figure*}

We now present results for mesh networks under single-node expansion.
We focus on comparisons between optimized and evolved networks,
as well as between random and gradual expansion.

The computational experiments are performed as follows.
We consider a rectangular area of length 3000 and width 1500 
(roughly the aspect ratio of the continental US) and 500 potential locations 
in that area. We start from a randomly chosen location
and in each step we expand the network (randomly or gradually)
by adding one more location. We design the optimized and evolved 
networks so that they always interconnect the same set of locations. 
The maximum network consists of 60 locations - this is a realistic scale for
the backbone of a service provider.
The experiments are repeated 20 times, and we
report 90\% confidence intervals for all results.

The delay bound is set to $D=1.3\times d$, where d is 
the length of the diagonal in the previous rectangle. 
With this value of the delay bound, the designed  networks
are sparse (the number of links is typically at most twice 
the number of nodes), which is also a characteristic of 
the physical-layer backbone connectivity in practice \cite{lunli:04}. 
Further, with this value of $D$
we can always compute an acceptable network
using the algorithms of the previous section.
 
We have also experimented with other delay bounds between 
$1.2 \times d$ and $2.5\times d$, without observing significant
qualitative differences. As $D$ increases, the resulting networks
get sparser and after a certain point they become rings.  
As $D$ approaches $d$, on the other hand, it becomes likely 
that there are no acceptable networks for a given set of locations
and the networks can be unrealistically dense. 

The costs of the optimized and evolved networks are shown in 
Figures~\ref{fig:basic r Costs} and ~\ref{fig:basic g Costs} 
for random and gradual expansion, respectively. 
We have confirmed that these costs scale as $\sqrt{n}$
in the case of random expansion, and as $n$ in the case
of gradual expansion (the regression lines are omitted for clarity).
Interestingly, these are the same scaling expressions we derived
in the case of ring networks. 

The modification costs also scale as ring networks,
even though there is significantly higher variability in mesh networks.
Specifically, the modification cost decreases as $1/\sqrt{n}$ under 
random expansion, and it remains practically constant under gradual expansion
(Figure~\ref{fig:basic rg Cmod}).
All previous costs are higher in random than in gradual expansion.
This is because gradual expansion leads to much shorter links
(Figure~\ref{fig:basic rg average linklength}).

In terms of cost overhead, the scaling analysis for rings predicts
that $v(n)$ should not depend on the network size for large values of $n$.
Figures~\ref{fig:basic r metrics} and \ref{fig:basic g metrics}
show the cost overhead for mesh networks under random and gradual
expansion, respectively.
The variability across different expreriments is large, and so
we use the non-parametric Mann-Kendall hypothesis test for
trend detection. Indeed, when we focus on the larger values of $n$, say
$n>20$, the test {\em cannot reject} the null hypothesis that the
cost overhead shows no trend (p-value = 0.36 for random expansion
and 0.34 for gradual expansion).  
So, we expect that the cost overhead does not increase under
single-node expansion, even in the case of mesh networks. 
Additionally, as expected from the case of rings, 
the cost overhead is significantly higher under random expansion than 
gradual expansion.

Figures~\ref{fig:basic r metrics} and ~\ref{fig:basic g metrics}
also show the evolvability under random and gradual expansion. 
They both increase fast until they become approximately equal to one.
So, as in the case of rings, it is less costly to incrementally
modify an existing network instead of re-designing it from scratch.
It is interesting that the evolvability under gradual expansion
is slightly less than under random expansion when $n<20$. 
This is a consequence of the lower optimized network cost under
gradual expansion - the modification cost is not significantly
lower than the optimized cost in that range of network sizes. 

The topological similarity between the corresponding optimized 
and evolved networks is shown in Figure~\ref{fig:basic rg T}. 
Even though it does not show a clear trend with $n$, 
it is interesting that the two networks share about 
50-70\% of their links, when $n>20$. 
Thus, the two network topologies
become (and stay) significantly different during the expansion
process (even though they started from the same three-node ring).

The inventory overhead is shown in Figure~\ref{fig:basic rg R}.
There is a decreasing trend under both random and gradual expansions.
The fact that the inventory overhead does {\em not} increase 
means that the inventory does not become increasingly
more costly relative to the cost of the evolved network. 
If that was the case, the evolved network would gradually 
accumulate a ``baggage'' of unused links with 
increasing cost relative to the cost of the network itself.
The opposite happens: even though the cost of 
the inventory increases in absolute terms, it decreases
relative to the cost of the network.

We have also compared the performance of the two networks,
in terms of the path propagation delay, when there are no failures.
In that case only the primary path is used for each pair of nodes. 
Figure~\ref{fig:basic rg pathlength} shows the ratio of
the propagation delay in the primary path between the optimized
and evolved networks.
The evolved network, under random expansion, gives significantly 
lower propagation delay than the corresponding optimized network.   
One reason is that the former has larger link density.
A second reason, related to the presence of hubs in the evolved
network, is discussed in Section~\ref{sec:robustness}. 
The difference is not significant under gradual expansion 
because the links in those networks are of comparable number and length.

Another significant difference between random and gradual expansion
is shown in Figure~\ref{fig:basic rg average linklength}. 
Under gradual expansion, the average link length remains practically
constant as the network grows.

Under random expansion, the average link length is consistently 
higher than under gradual expansion, but it decreases with $n$.
The reason is that, as the network grows, it gradually covers 
a larger span of the rectangular region in which all potential nodes 
are located. So, the need for longer links is gradually decreased
and the new links that are added in each environment get shorter over time. 

\section{Multi-Node Mesh Expansion}
	\label{sec:rho-effect}
In this section, we consider multi-node expansion in mesh networks. 
An expansion factor $\rho$ means that the
network expands at a single environment 
from an initial size of $n$ nodes to $\lfloor \rho \, n \rfloor$ nodes. 
In the following experiments, $n$=15 nodes.

\begin{figure}[ht]
\begin{center}
\subfigure[Evolvability]{
\scalebox{0.5}{\input{Figures/pace/ERG.tex}}
\label{fig:pace ERG}
}\\
\subfigure[Cost overhead]{
\scalebox{0.5}{\input{Figures/pace/VRG.tex}}
\label{fig:pace VRG}
}
\caption{Effect of expansion factor $\rho$ on (a) evolvability and  (b) cost overhead under random and gradual multi-node expansion.}
\end{center}
\label{fig:pace RG}
\end{figure}

\begin{figure*}[ht]
\begin{center}
\begin{tabular}[t]{cccc}
\subfigure[Empirical CDF]{
\includegraphics[angle=270, width=0.3\textwidth]{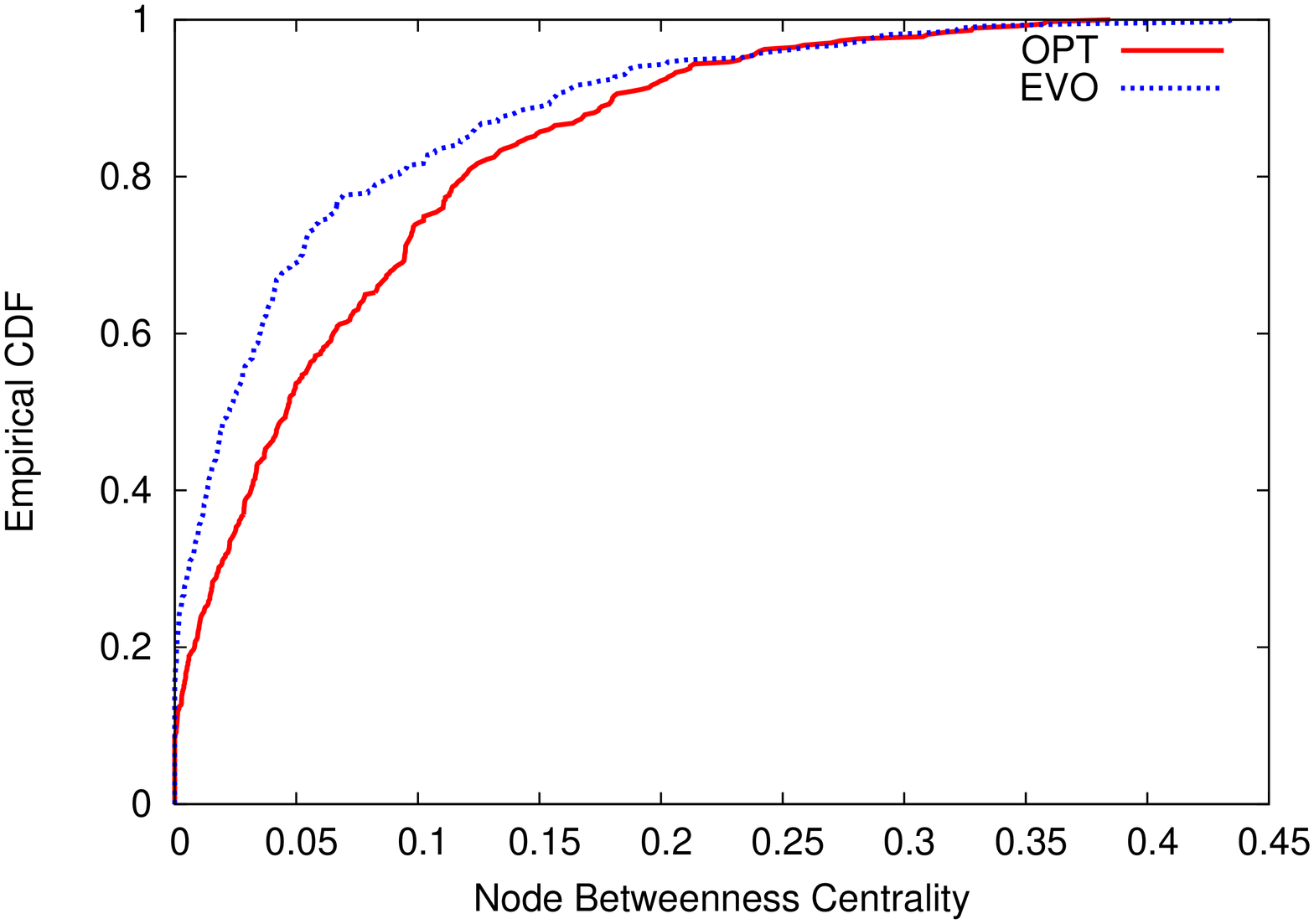}
\label{fig:robustness cdf}
}
\subfigure[Evolved Network]{
 \includegraphics[angle=270, width=0.3\textwidth]{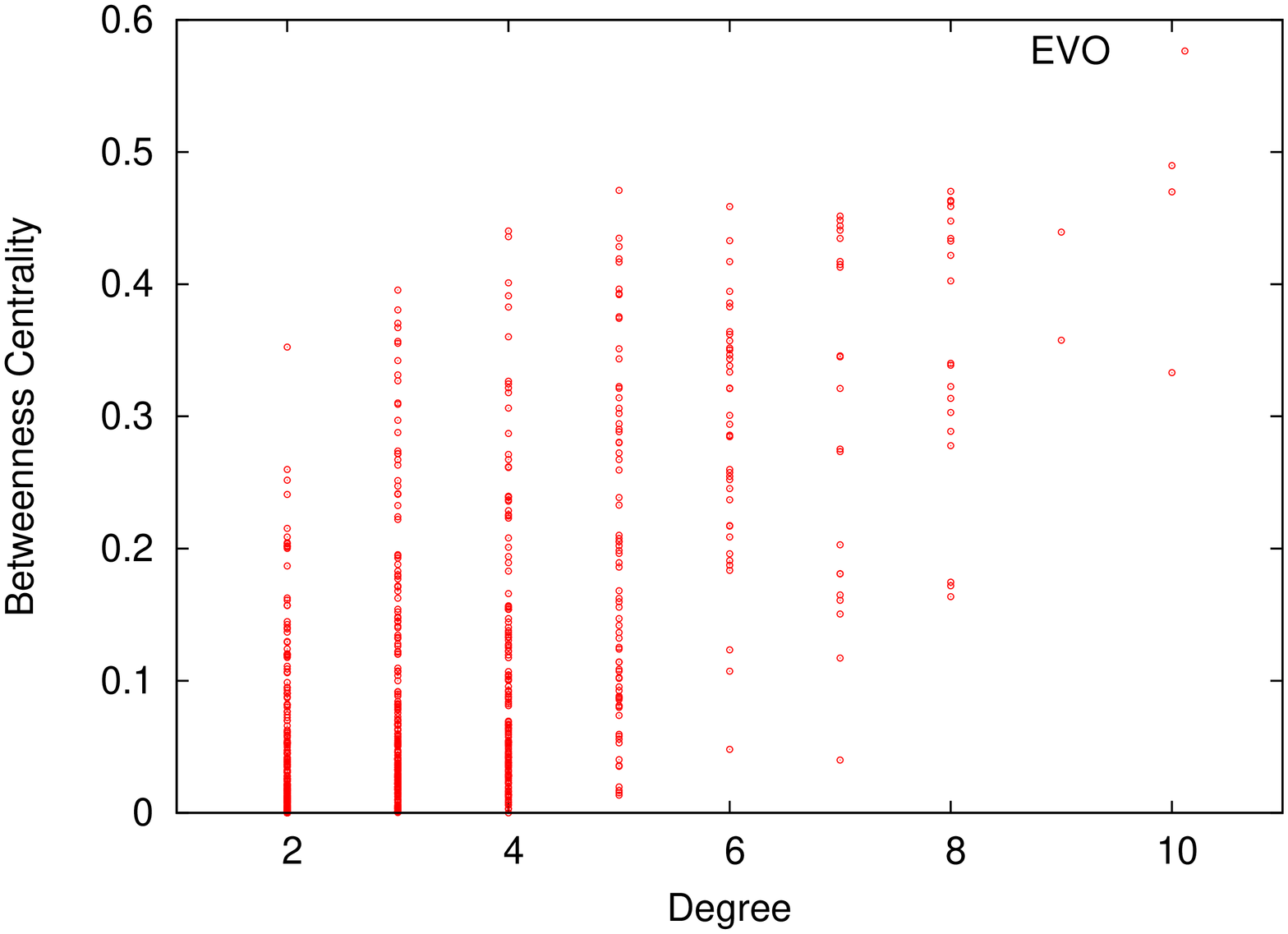}
\label{fig:robustness evo}
}
\subfigure[Optimized Network]{
 \includegraphics[angle=270, width=0.3\textwidth]{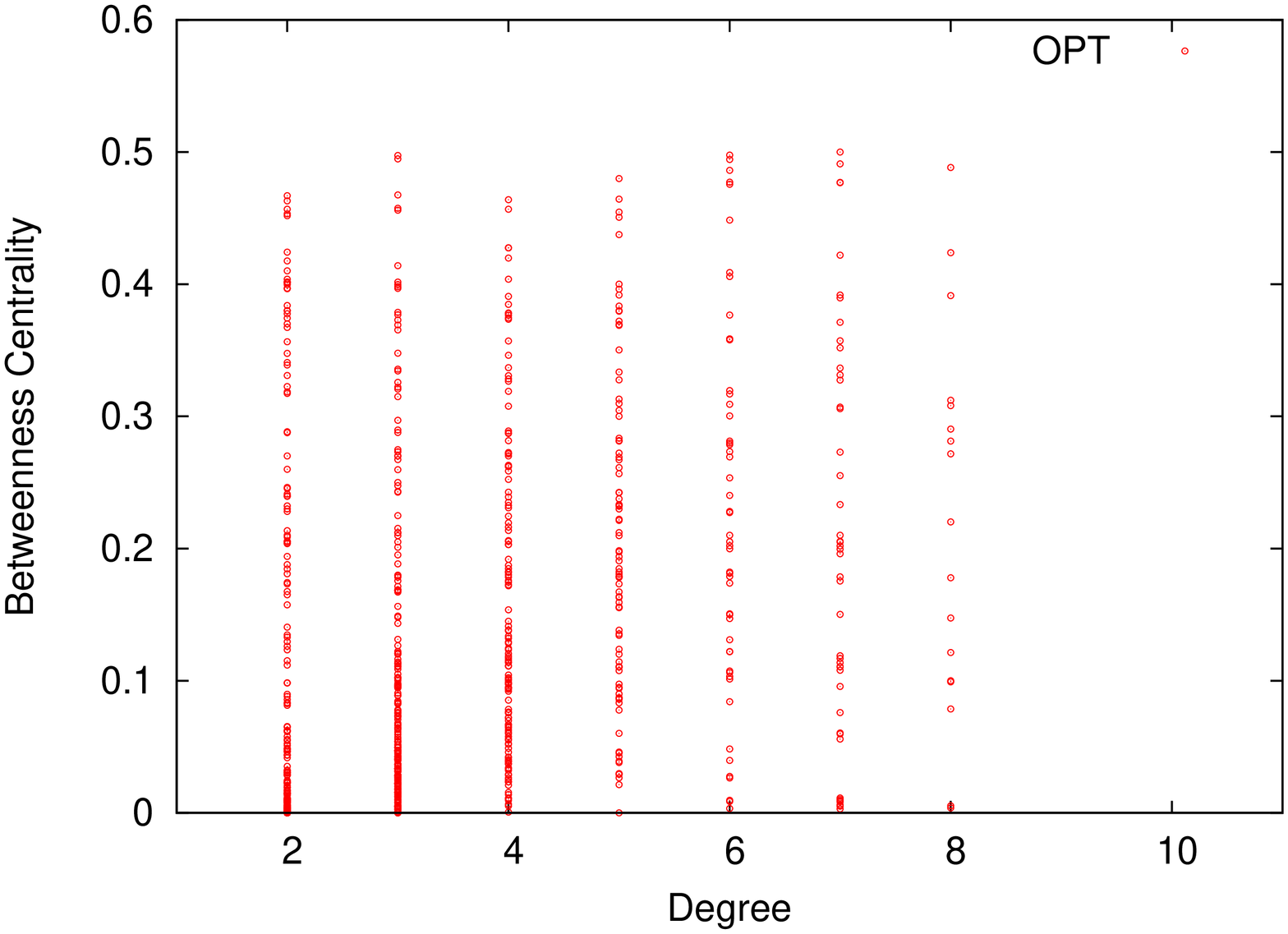}
\label{fig:robustness opt}
}
\end{tabular}
\caption{Robustness graphs under single-node random expansion(n=50)}
\label{robustness}
\end{center}
\end{figure*}

The initial evolved network (before the multi-node expansion)
is designed using single-node expansion, until it reaches size $n$. 
We focus on the effect of $\rho$ on the two main metrics: 
the evolvability $e(\rho)$ and cost overhead $v(\rho)$.

Figure~\ref{fig:pace ERG} shows the evolvability under random and 
gradual expansion as $\rho$ increases.
As in the case of rings, the evolvability decreases with $\rho$,
and the decrease is faster under random expansion.
The critical expansion factor $\hat{\rho}$ under random expansion 
is larger than four. 
Thus, at least in these computational results, 
it is less costly to modify the 
existing network incrementally than to redesign it from scratch
if the network size increases by less than a factor of four.
On the other hand, the evolvability under gradual expansion
remains positive in the range of network sizes that we could design
(as in the case of rings). 
It is an open question whether the evolvability can ever be negative
under gradual expansion. 

Figure~\ref{fig:pace VRG} shows that the cost overhead
increases with $\rho$ under both random 
and gradual expansion. The latter leads to significantly 
lower values, as in the case of rings. 
The increase of the cost overhead, however, is concave under
both random and gradual expansion, and it does not exceed 100\%,
at least in these computation results, when $\rho$ is less than four. 

The scaling expressions that were derived for ring networks 
assuming that $\rho$ is close to one
(see Equations~\ref{eq:rnd-evo-rho}, \ref{eq:grd-evo-rho},
\ref{eq:rnd-ovh-rho} and \ref{eq:grd-ovh-rho})
also give accurate regression curves for mesh networks
(at least when $\rho < 3.5$).
Recall however that these expressions should not be used to examine
the asymptotic behavior of the evolvability or cost overhead 
as $\rho$ increases. 

\section{Centrality and Robustness}
	\label{sec:robustness}
The generated networks are robust to single node or link failure 
because they have two node-disjoint paths (primary and secondary)
between each pair of nodes.
We could compare the robustness of the designed topologies by
considering multiple link or node failures that are randomly generated
in a simulator. 
We rely instead on a more abstract network analysis approach based 
on the {\em betweenness centrality} metric. 
Specifically, we define as Betweenness Centrality of a node (node-BC) 
the fraction of primary paths that traverse that node, among all 
primary paths in the network.
Similarly, the Betweenness Centrality of a link (link-BC) is
the fraction of primary paths that traverse that link.
The nodes (or links) with the highest BC values
can be thought of as the network's most critical 
components; if they are somehow perturbed (without necessarily failing), 
the impact on the entire network will be much higher than if we
perturb nodes (links) with low BC.
Similarly, we can compare the robustness of two networks X and Y 
that have the same number of nodes (and thus the same number of 
primary paths) using the BC metric.
If the average node-BC across all nodes in X is higher than in Y, 
network X is more susceptible (or less robust) to node perturbations 
than network Y; similarly for link perturbations.

\begin{figure}[ht]
\begin{center}
\includegraphics[width=0.3\textwidth]{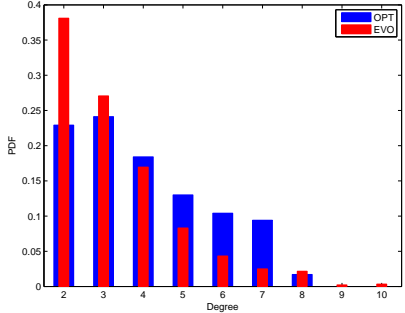}
\label{fig:robustness degs}
\caption{Degree distribution of evolved and optimized networks.}
\label{robustness}
\end{center}
\end{figure}
We compare next the robustness of evolved and optimized networks 
under single-node random expansion. 
Due to space constraints we focus on the node-BC metric; the results are
similar for link-BC. 
Figure~\ref{fig:robustness cdf} shows the empirical CDF of the BC 
across all nodes in networks of size $n$=50. 
These empirical CDFs are constructed from 20 independently
generated networks
(i.e., the sample size in each empirical CDF is 1000 values).  
Note that the nodes of the optimized network have
higher betweenness centrality than the nodes of the evolved network, 
at least when the BC is less than 20\%. 
The average node-BC is 0.15 in the optimized and 0.10 in the evolved networks.
So, we expect that the evolved network will be more robust
to node perturbations, on average, than the optimized network. 

There is an interesting difference between the two networks however,
which is not evident from the previous CDFs.
Figure~\ref{fig:robustness evo} shows a scatter plot for the node degree 
and BC in the 20 evolved networks we consider,
while Figure~\ref{fig:robustness opt} shows the corresponding 
scatter plot for the 20 optimized networks. 
In the evolved networks, there is a strong positive
correlation (Pearson's correlation coefficient: 84\%) between
node degree and BC, 
while there is no significant correlation in optimized networks 
(Pearson's correlation coefficient: 11\%).  
In other words, in the evolved network the most critical nodes
(highest BC) are also the nodes with the largest number of connections;
this is not the case in optimized networks.

We also compare the degree distribution of evolved and optimized networks
(see Figure~\ref{robustness}).
In evolved networks, about 4\% of the nodes have a degree of 
eight or more; the corresponding percentage is 2\% in optimized
networks and there are no nodes with degree higher than eight.  
Additionally, the percentage of nodes with only two links
(the minimum degree that is necessary to satisfy the reliability constraint)
is 38\% in evolved and 23\% in optimized networks. 
In other words, the evolved networks also have a more skewed degree
distribution (the skewness of the degree distribution is 1.2 in optimized 
and 1.5 in evolved networks).
We refer to those higher-degree nodes as {\em hubs.} 
It is the hubs that also have the highest BC in evolved networks.

Why is it that the incremental design process creates nodes with 
considerably larger degree and BC than most other nodes in the same network? 
The evolved network at environment $k$ is generated from 
the corresponding network at environment $k-1$; so, any existing hubs
are inherited to the network of the next environment. 
Further, a hub offers short paths to many other nodes (due to its 
large number of connections). So, if the addition of a new node in 
the network causes a violation of the reliability or delay constraints, 
it is more likely that a new hub connection will resolve that violation
than a new connection to a low degree node.
We have also confirmed that the nodes with the highest degree in 
evolved networks are nodes that were created early in the incremental
design process, accumulating a large degree over time (results not
shown due to space constraints). 
On the contrary, the optimized design process creates a new network
at each environment, and so it is not likely that a hub at environment
$k-1$ will also be a hub at environment $k$. 

In summary, evolved networks are more robust than optimized networks,
in terms of the average BC metric we consider. If we randomly
select a node in each network and perturb it, we expect that the impact
of the perturbation will be higher in the optimized network. 
At the same time however, an evolved network  has a small number of 
hubs that also have high BC, relative to the rest of the nodes in that 
network; those nodes represent ``Achilles' heel'' in evolved networks.  
Further, evolved networks have more hubs (more nodes at the
tail of the degree distribution) than optimized networks, and so 
evolved networks are more susceptible to hub perturbations 
than optimized networks. 
These observations are similar to a well-known finding about 
scale-free networks \cite{attack-tolerance}: those networks are robust
to the failure of randomly selected nodes but are fragile to the failure
of hubs. We cannot claim, however, that the evolved networks
we consider in this paper are scale-free because of the limited network 
sizes we can generate computationally. The connection between our study
and the literature of scale-free networks is an interesting problem for
future research.

\section{Ownership and Leasing}
	\label{sec:inventory}
In the previous mesh network design sections, 
we adopted the {\em Inventory} option, in which any unnecessary links
are ``turned off'' and they are (physically or virtually) moved to an 
inventory so that they can be reused in the future if needed.
As discussed in Section~\ref{sec:framework}, 
we can also consider two variations of the incremental design process  
that do not require an inventory: an {\em Ownership} option in which
the network maintains existing links even if they are not necessary,
and a {\em Leasing} option in which links that are not necessary 
in the current environment are removed (if those links are needed
again in the future, they will have to be re-purchased). 
In this section, we compare the Inventory option with the Ownership
and Leasing options under single-node expansion. 
Due to space constraints, we only summarize
the results without including graphs.

The Ownership option results in significantly higher
evolved network cost than the two other options 
(by a factor of about two in our random expansion experiments).
The reason is that the former leads to a much larger number of links.
What is more interesting, however, is that the difference between
the Inventory and Leasing options is statistically insignificant 
($C_{evo}$ is only marginally higher in the former).
In other words, even though the Inventory option keeps some 
unnecessary links in the network, those links are only few relative
to the total number of links in the network. This is related to
the earlier observation regarding the decreasing trend of the 
inventory overhead (Figure~\ref{fig:basic rg R}).

On the other hand, the modification cost is typically
higher in Leasing than in the two other options.
With Leasing, the incremental design process has to occasionally 
acquire more new links than with the two other option.
This difference in $C_{mod}$, however, is 
insignificant compared to the optimized network cost 
$C_{opt},$\footnote{Obviously, $C_{opt}$ does not depend on
the three inventory options of the incremental design process.} 
and so the evolvability of the three options is 
practically the same (but with higher variance in the leasing option).

In terms of cost overhead, the Ownership option 
results in significantly higher values than the
two other inventory options (on the average about 0.6 versus
0.1, for large networks). The difference between 
Leasing and Inventory is not statistically significant at most environments. 

In summary, the incremental design process 
clearly benefits if it does {\em not} use the Ownership option.
The  Inventory option represents a good compromise
between minimizing $C_{evo}$ and $C_{mod}$.
When it is not possible, however, to maintain an 
inventory, the results of this section show that the 
Leasing option would result in similar evolvability 
and cost overhead with the Inventory option.

\section{Related Work}
	\label{sec:relatedwork}
The topology design literature is extensive both in 
the domain of computer networks and in theoretical computer science,
and it is well covered in a recent book 
by Pioro and Medhi \cite{network-design}.
The vast majority of that literature, however, focuses on 
optimized network design.
Only few studies have focused on incremental network design
(also referred to as ``multi-period design''), 
and none of them, to the extent of our knowledge,
have focused on a comparison between incremental and optimized design.

Specifically, there are some studies that focus on 
incremental network design 
\cite{incremental,survivable,optic,network,distribution-network,Wu,Geary01,Pickavet,Yaged73,Zadeh74} (see also section 11.2 of \cite{network-design}).
Those works mostly propose algorithms and optimization frameworks for 
incremental network design under a wide range of different 
constraints and objectives.
Some of them consider topology design while others consider 
capacity expansion coupled with routing changes, and some of
them consider reliability constraints while others consider 
limited budget constraints. 
None of them, however. is significantly relevant to our study
because they do not compare incremental designs with the 
corresponding optimized designs, and they do not consider 
different expansion models (e.g., random versus gradual
or single-node versus multi-node). 

Chiang and Yang have focused on various ``X-ities", such as evolvability, 
scalability, reliability, or adaptability in the context of computer 
networks \cite{chiang2004tnx}.
They present an analytic framework to capture the notions of 
evolvability and scalability. They also present an Evolvable Network Design 
(END) Tool using dynamic programming methods to design the multi-phase 
deployment of a network so that early phase designs are more evolvable 
in later stages. 

A quite different, but still relevant, study by Tero et al.
\cite{AtsushiTero01222010} compared the Tokyo rail system 
(as an example of an optimally designed transportation network)
with a natural network formed by the slime mold Physarum polycephalum.
The slime mold was allowed to grow on a rectangular map of the city
of Tokyo; the map contained food on the locations at which the 
Tokyo rail system has stations. 
The slime mold network grew in an incremental manner, without any 
centralized control or ``intelligence'', and it gradually 
covered and interconnected all food locations while refining its
connectivity over time.
The authors compared the two networks in terms 
of efficiency, fault tolerance, and cost and
found that they are actually quite similar! 
This novel experiment implies that even a simple and incremental 
design process may be able to produce  a network that has the 
cost efficiency and reliability properties that we usually only
expect from optimized networks. 

\section{Conclusions}
	\label{sec:conclusions}
We can now return to the questions that were asked in  
the introduction and summarize our main findings.
The following conclusions are supported by asymptotic 
scaling expressions for rings and by computational  
results in the case of mesh networks; it appears though
that the analytical expressions for rings also fit well the computational
results for the mesh networks we experimented with.\\
\noindent
1. We formulated the incremental network design process as
an optimization problem that aims to minimize the modification
cost relative to the previous network.
We also identified and compared certain expansion models 
(random versus gradual, and single-node versus multi-node) 
and three variations of the incremental
design process (Inventory, Ownership, Leasing). \\
2. Even though an evolved network has higher
cost than the corresponding optimized network, 
the cost overhead of the former does not increase as the network
grows, at least under single-node expansion.\\
3. In the case of mesh networks,
the incremental design process leads to networks 
with larger link density, higher performance in terms of the
average propagation delay across all primary paths, and improved
robustness in terms of a node betweenness centrality metric, compared
to optimized networks. 
These differences are more pronounced under random expansion.\\ 
4. Under single-node (random or gradual) expansion, 
it is less costly to follow the incremental design approach 
than to re-design the network from scratch. 
The evolvability under basic expansion approaches one as 
the network grows. \\
5. Under multi-node and random expansion, there is a critical value $\hat{\rho}$
of the expansion factor beyond which it is less costly to abandon the existing
network and re-design the network from scratch. It is not clear
whether this is ever the case under gradual expansion; our computational
experiments have never produced negative evolvability in that case. \\
6. The incremental and optimized design processes lead to significantly
different network topologies.
The evolved network has a more skewed degree distribution compared
to the optimized network, and it includes few nodes (hubs) with much 
higher degree and betweenness centrality than most other nodes.\\ 
7. The inventory overhead of the incremental design process does
not increase with time, and so the cumulative cost of the
inventory does not diverge relative to the cost of the evolved network.\\ 
8. Under gradual expansion, the evolvability is higher
and the cost overhead is lower than under random expansion.
The model of  gradual expansion represents a more ``evolution-friendly''
dynamic environment than random expansion.\\  
9. The Inventory option is a good compromise between cost overhead
and modification cost, compared to the Ownership and Leasing options.
If it is not possible to maintain an inventory, the Leasing option 
performs quite similar to the Inventory option.

In terms of future work, we believe that the previous questions can be 
studied more mathematically for specific regular or random network topologies,
deriving exact expressions. 
It would also be interesting to examine other dynamic environment models,
such as iterated multi-node expansion, models in which the traffic
loads and link capacities change with time as well, as well as more 
elaborate economic models that involve discounting or dynamic costs.
It would also be interesting to see this quantitative framework
and comparisons between evolved and optimized designs applied in other   
problems and technological domains.
\bibliography{papers}

\end{document}